\documentclass[aps,preprint,nofootinbib,superscriptaddress]{revtex4}

\usepackage[english]{babel}
\usepackage[latin1]{inputenc}
\usepackage{graphicx}
\usepackage{amsmath}
\usepackage{color}
\usepackage{bm}
\usepackage{ulem}

\def\d{\mathrm{d}}

\def\pp{\mathbf{p}}

\def\kk{\mathbf{k}}

\def\rr{\mathbf{r}}

\def\QQ{\mathbf{Q}}

\newcommand{\mrm}[1]{\mathrm{#1}}
\newcommand{\mbf}[1]{\mathbf{#1}}

\def\Ordo{\mathcal{O}}
\def\Lag{\mathcal{L}}
\def\beq{\begin{equation}}
\def\eeq{\end{equation}}

\def\bdm{\begin{displaymath}}
\def\edm{\end{displaymath}}

\newcommand{\mR}[0]{m_\mrm{R}}
\newcommand{\kC}[0]{k_\mrm{C}}

\newcommand{\Zc}[0]{Z_\mrm{c}}

\def\bea{\begin{eqnarray}}
\def\eea{\end{eqnarray}}

\newcommand{\nuc}[2]{\ensuremath{{}^{#1}\mathrm{#2}}}

\begin{document}

\title{Constraining Low-Energy Proton Capture on Beryllium-7
through Charge Radius Measurements}
\author{Emil Ryberg}
\affiliation{Department of Fundamental Physics, Chalmers University of
  Technology, 41296 Gothenburg, Sweden}
\author{Christian Forss\'en}
\affiliation{Department of Fundamental Physics, Chalmers University of
  Technology, 41296 Gothenburg, Sweden}
\author{H.-W. Hammer}
\affiliation{Institut für Kernphysik, Technische Universität Darmstadt, 
64289 Darmstadt, Germany}
\affiliation{ExtreMe Matter Institute EMMI, GSI Helmholtzzentrum 
für Schwerionenforschung GmbH, 64291 Darmstadt, Germany}
\author{Lucas Platter}
\affiliation{Argonne National Laboratory, Physics Division, Argonne, IL 60439, USA}
\affiliation{Department of Fundamental Physics, Chalmers University of
  Technology, 41296 Gothenburg, Sweden}
\date{\today}

\begin{abstract}
  In this paper, we point out that a measurement of the
  charge radius of Boron-8 provides indirect access to the S-factor
  for radiative proton capture on Beryllium-7 at low energies. 
  We use leading-order
  halo effective field theory to explore this correlation and we give a
  relation between the charge radius and the S-factor. Furthermore, we
  present important technical aspects relevant to the renormalization
  of pointlike P-wave interactions in the presence of a repulsive
  Coulomb interaction.
\end{abstract}

\smallskip
\maketitle
\newpage
\section{Introduction}
\label{sec:introduction}
A frequent problem encountered in nuclear astrophysics is that
observables relevant to stellar and cosmological evolution cannot be
measured directly in the laboratory. In many cases, this is due to the
Coulomb barrier, which leads to very small cross sections at the
low energies at which these processes occur in the stellar
environment. Specifically, electroweak reactions relevant to big bang
nucleosynthesis or solar astrophysics that involve either neutrinos or
soft photons display this problem. A common strategy is then to carry out 
experiments of the desired observable at
higher energies and to use models to extrapolate to the low-energy 
region. One process that has
been analyzed frequently this way is radiative proton capture on \nuc{7}{Be}
into \nuc{8}{B}. In the solar $pp$ fusion reaction network, this reaction
proceeds at $\sim$10~keV energies, and the subsequent beta decay 
determines the intensity of the high-energy part of the solar neutrino spectrum. 
The currently accepted threshold value of the corresponding S-factor 
\begin{equation}
S(0)=(20.8\pm 0.7 ({\rm expt.})\pm 1.4 ({\rm theor.}))~\mrm{eV~b}
\label{eq:recvalS}
\end{equation}
has been obtained by extrapolation of experimental data 
\cite{Adelberger:2010qa}.

Effective field theory (EFT) provides an alternative way to obtain
information on such observables. The EFT algorithm correlates a finite
number of physical observables at each order, based on a power
counting scheme that captures the relevant scales of the system. The
simplest example might be the correlation between the scattering
length $a$ and the binding energy of two particles with mass $m$,
$B_2=1/(ma^2)$, when $a$ is positive and much larger than the range of
the interaction $R$.  The application of these ideas to nuclear
physics has led to the development of the so-called pionless EFT,
which is applicable for typical momenta below the pion mass.  (See
Refs.~\cite{Beane:2000fx,Bedaque:2002mn,Epelbaum:2008ga} for reviews
and references to earlier work.) If one also introduces field
operators for certain tightly bound clusters of nucleons, such as
$\alpha$ particles, the pionless EFT is usually referred to as halo
EFT \cite{Bertulani:2002sz,Bedaque:2003wa}. It has been applied to
describe the structure and reactions of a number of known and
suspected one- and two-neutron halo systems in the Helium
\cite{Rotureau:2012yu,Ji:2014wta}, Lithium
\cite{Canham:2008jd,Rupak:2011nk,Fernando:2011ts,Zhang:2013kja,Hagen:2013xga},
Beryllium \cite{Canham:2008jd,Hammer:2011ye,Hagen:2013xga}, Carbon
\cite{Canham:2008jd,Rupak:2012cr,Acharya:2013nia,Acharya:2013aea,Hagen:2013xga},
and Calcium isotope chains \cite{Hagen:2013jqa}.

For reactions with two or more charged particles, Coulomb effects are
important.  Although the Coulomb interaction is perturbative for
intermediate and higher energies, it has to be treated
nonperturbatively close to threshold.  In the two-nucleon sector this
was first discussed by Kong and Ravndal for the proton--proton
channel~\cite{Kong:1998sx,Kong:1999sf} and later extended to
next-to-next-to-leading order by Ando {\it et al.}~\cite{Ando:2007fh}.
A renormalization-group analysis of proton--proton scattering in a
distorted-wave basis was performed in
Refs.~\cite{Barford:2002je,Ando:2008jb} and $\alpha\alpha$ scattering,
which is governed by a shallow S-wave resonance close to threshold,
was discussed in Ref.~\cite{Higa:2008dn}.  The extension to describe
the structure and charge radii of S-wave proton halo bound states with
repulsive Coulomb interaction was recently carried out in
Ref.~\cite{Ryberg:2013iga}.

Here we focus on \nuc{8}{B} as a loosely bound state of
\nuc{7}{Be}-proton. This system has been
considered previously in halo EFT by Zhang, Nollett and
Phillips~\cite{Zhang:2014zsa}.  They calculated the S-factor of the
reaction \nuc{7}{Be}$(p,\gamma)$\nuc{8}{B} at leading order (LO) and
obtained $S(0)=(18.2\pm 1.2 ({\rm ANC\ only}))~\mrm{eV~b}$ consistent
with the recommended value in Eq.~(\ref{eq:recvalS}). Moreover, they
performed a detailed error analysis and investigated the sensitivity
of their result to the input parameters.  The scope of our work is
somewhat different.  Modern isotope shift measurements facilitate an
extremely precise determination of the nuclear charge radius.  We will
show below that the charge radius of \nuc{8}{B} and the astrophysical
S-factor of the reaction \nuc{7}{Be}$(p,\gamma)$\nuc{8}{B} are free of
short-distance counterterms at LO in halo EFT. Thus, these two
observables are correlated at this order and a measurement of the
charge radius can provide independent information on the S-factor.

Our strategy is twofold: First, we match our low-energy coupling
constants to the proton separation energy of 8B and the asymptotic
normalization coefficients (ANCs) determined from either ab initio
nuclear-structure calculations or from transfer reaction measurements
\cite{Nollett:2011qf,Navratil:2011sa,Tabacaru:2005hv} in order to
calculate both the charge radius of $^8$B and the S-factor at
threshold. Second, we demonstrate the leading-order correlation
between these two observables, and we show explicitly how a datum for
the charge radius of $^8$B would put a constraint on the threshold
astrophysical S-factor for $^7$Be(p, $\gamma$)$^8$B. In this connection, we
employ two different field theories; with and without explicit
inclusion of the excited-core state. The incorporation of the
excited-core field corresponds to an inclusion of additional
short-distance physics, and therefore allows to study the stability of
the EFT approach in LO calculations.

The paper is structured as follows. In
Sec. \ref{sec:lagr-feynm-rules}, we introduce the halo EFT for the $^
8$B system. We provide a detailed discussion of the renormalization in
a P-wave channel with Coulomb interactions and the relevant low-energy
constants at LO.  In Sec. \ref{sec:electric-form-factor}, we calculate
the charge form factor of \nuc{8}{B}. The radiative capture reaction,
\nuc{7}{Be}$(p,\gamma)$\nuc{8}{B}, is considered in
Sec.~\ref{sec:radiative-capture}, and the correlation between the
cross section at threshold and the charge radius of \nuc{8}{B} is
discussed in Sec.~\ref{sec:corr-charge-radi}. We conclude with a
summary and an outlook in Sec. \ref{sec:summary}. Some properties of the
Coulomb Green's function and P-wave integrals are reviewed in the Appendices.
\section{Halo EFT for P-wave interactions between charged particles}
\label{sec:lagr-feynm-rules}
The $J^\pi = 2^+$ ground state of $\nuc{8}{B}$ can be viewed as a halo
state consisting of a ($J^\pi = 3/2^-$)
$\nuc{7}{Be}$ core and a proton that is bound in a relative P-wave. 
This system is amenable to Halo EFT since it displays a
separation of scales. The one-proton separation energy of $\nuc{8}{B}$ is 
$B=0.1375~\mrm{MeV}$ and the energy of the first excited state
of the $\nuc{7}{Be}$ core is $E^*=0.4291$~MeV. In our EFT approach, we will
include the proton, the ground-state
$\nuc{7}{Be}$-core and its first excited state as explicit degrees of 
freedom. An estimate for the  breakdown scale of this EFT is given by
the lowest state not included within this theory. Since the energy of the 
second excited state of $\nuc{7}{Be}$ is relatively high, at 4.57~MeV,
the relevant high-energy breakdown scale is rather the threshold for
breakup of $\nuc{7}{Be}$ into $\nuc{3}{He}$ and  $\nuc{4}{He}$ at
$E_\alpha = 1.5866$~MeV~\cite{Tilley:2002aa}.

The Lagrangian for this system, including the P-wave interaction between the 
proton and the $\nuc{7}{Be}$ core ground and excited state, is given 
by \cite{Zhang:2014zsa}
\begin{align}
\label{eq:lagrangian}
\Lag=&
{\rm p}_\sigma^\dagger\Big(i\mrm{D}_t+\frac{\mbf{D}^2}{2m}\Big){\rm p}_\sigma
+\mrm{c}^\dagger_a\Big(i\mrm{D}_t+\frac{\mbf{D}^2}{2M}\Big)\mrm{c}_a\nonumber\\
&+\mrm{\tilde{c}}^\dagger_\sigma\Big(i\mrm{D}_t+\frac{\mbf{D}^2}{2M}-E^*\Big)\mrm{\tilde{c}}_\sigma
+\mrm{d}_\alpha^\dagger\Big[\Delta+\nu\Big(i\mrm{D}_t+\frac{\mbf{D}^2}{2M_\mrm{tot}}\Big)\Big]\mrm{d}_\alpha\nonumber\\
&-g_1\Big[\mrm{d}^{\dagger}_\alpha \mathcal{C}^\alpha_{j k} \mathcal{C}^j_{a\sigma}\mathrm{c}_a
\left((1-f)i\overrightarrow{\nabla}_k-fi\overleftarrow{\nabla}_k\right)\mrm{p}_\sigma+\mrm{h.c.}\Big]\nonumber\\
&-g_2\Big[\mrm{d}^{\dagger}_\alpha \mathcal{C}^\alpha_{\beta k} \mathcal{C}^\beta_{a\sigma}\mathrm{c}_a
\left((1-f)i\overrightarrow{\nabla}_k-fi\overleftarrow{\nabla}_k\right)\mrm{p}_\sigma+\mrm{h.c.}\Big]\nonumber\\
&-g_*\Big[\mrm{d}^{\dagger}_\alpha \mathcal{C}^\alpha_{j k} \mathcal{C}^j_{\sigma\chi}\mathrm{\tilde{c}}_{\chi}
\left((1-f)i\overrightarrow{\nabla}_k-fi\overleftarrow{\nabla}_k\right)\mrm{p}_\sigma+\mrm{h.c.}\Big]+\ldots~,
\end{align}
where $\mathrm{p}^\sigma$ denotes the spin-1/2 proton field with mass
$m$, $\mathrm{c}^a$ ($\mathrm{\tilde{c}}^a$) the $\nuc{7}{Be}$ core
with spin-3/2
(spin-1/2 excited state) field with mass $M$. 
The parameter $f$ denotes
the mass ratio $m/M_{\rm tot}$, where $M_{\rm tot}=M+m$. The covariant
derivatives are defined as $\mrm{D}_\mu=\partial_\mu+i e \hat{\mrm{Q}}
A_\mu$, where $\hat{\mrm{Q}}$ is the charge operator. The halo field
$d_\alpha$ describing the $\nuc{8}{B}$ halo state
has mass $M_\mrm{tot}$, while the residual mass $\Delta$ is
needed for renormalization. The signature $\nu=\pm1$ will be chosen 
to reproduce the correct effective range \cite{Bertulani:2002sz}. 

The vertices where the halo field breaks up into a proton and a core
have strengths $g_1$ and $g_2$ for the $S=1,2$ spin-channels,
respectively. Furthermore, the vertex where the halo field breaks up
into a proton and an excited core in the $S=1$ channel has strength
$g_*$. We define the gound-state core coupling $g^2=g_1^2+g_2^2$,
since the $S=1,2$ channels will always appear together in this
combination for the observables that we consider. At LO, we have
therefore the three independent coupling constants $\Delta$, $g$ and
$g_*$, whose values will be determined by the proton separation energy of $^8$B
and two additional pieces of information on the $^7$Be core-proton and $^7$Be
excited-state-proton systems. The Clebsch-Gordan coefficients in
Eq.~(\ref{eq:lagrangian}) are defined as $\mathcal{C}^\alpha_{\beta
  k}=\left(2\beta1k\Big|\left(21\right)2\alpha\right)$,
$\mathcal{C}^\alpha_{jk}=\left(1j1k\Big|\left(11\right)2\alpha\right)$,
$\mathcal{C}^\beta_{a\sigma}=\left(\frac{3}{2}a\frac{1}{2}\sigma\Big|\left(\frac{3}{2}\frac{1}{2}\right)2\beta\right)$,
$\mathcal{C}^j_{a\sigma}=\left(\frac{3}{2}a\frac{1}{2}\sigma\Big|\left(\frac{3}{2}\frac{1}{2}\right)1j\right)$
and
$\mathcal{C}^j_{\sigma\chi}=\left(\frac{1}{2}\sigma\frac{1}{2}\chi\Big|\left(\frac{1}{2}\frac{1}{2}\right)1j\right)$. The
combinations $\mathcal{C}^{\alpha}_{jk}\mathcal{C}^{j}_{a\sigma}$,
$\mathcal{C}^{\alpha}_{\beta k}\mathcal{C}^{\beta}_{a\sigma}$ and
$\mathcal{C}^{\alpha}_{jk}\mathcal{C}^{j}_{\sigma\chi}$, where
repeated indices imply summation, give the two spin channels $S=1,2$
with the ground-state core and the $S=1$ channel for the excited core
state. The gradient $\nabla_k$ in the P-wave interaction vertex is
understood to be written as a spherical tensor, that is
$\nabla_{-1}=(\nabla_x-i\nabla_y)/\sqrt{2}$, $\nabla_0=\nabla_z$ and
$\nabla_{1}=(\nabla_x+i\nabla_y)/\sqrt{2}$. The possible values of the
spin indices are: $\alpha,\beta=-2,-1,0,1,2$, $a={\textstyle
  -\frac{3}{2},-\frac{1}{2},\frac{1}{2},\frac{3}{2}}$, $i,j,k=-1,0,1$
and $\sigma,\chi={\textstyle-\frac{1}{2},\frac{1}{2}}$. The terms
included in the Lagrangian above are the relevant pieces to describe
all observables at leading order in the low-energy expansion while the
ellipsis in Eq.~\eqref{eq:lagrangian} denote the operators required to
go beyond leading order.

An important ingredient of our approach is that we include the Coulomb
interaction between the proton and the core to all orders by using the
Coulomb Green's function $G_\mathrm{C}$. In terms of Coulomb
wavefunctions $\psi_\pp(\rr)$, see Appendix~\ref{app:Coulomb}, we can
write $G_\mrm{C}$ in coordinate space as
\begin{equation}
(\rr_1|G_\mrm{C}(E)|\rr_2)=
\int\frac{\d^3p}{(2\pi)^3}\frac{\psi_\pp(\rr_1)\psi^*_\pp(\rr_2)}{E-\pp^2/(2\mR)+i\varepsilon}~,
\label{eq:CGFa}
\end{equation}
where $\mR=mM/M_\mrm{tot}$ is the reduced mass.
Note that we are using round brackets to indicate coordinate-space
states. 

\subsection{The irreducible self-energy}

\begin{figure}[t]
\centerline{
\includegraphics*[scale=0.5,clip=true]{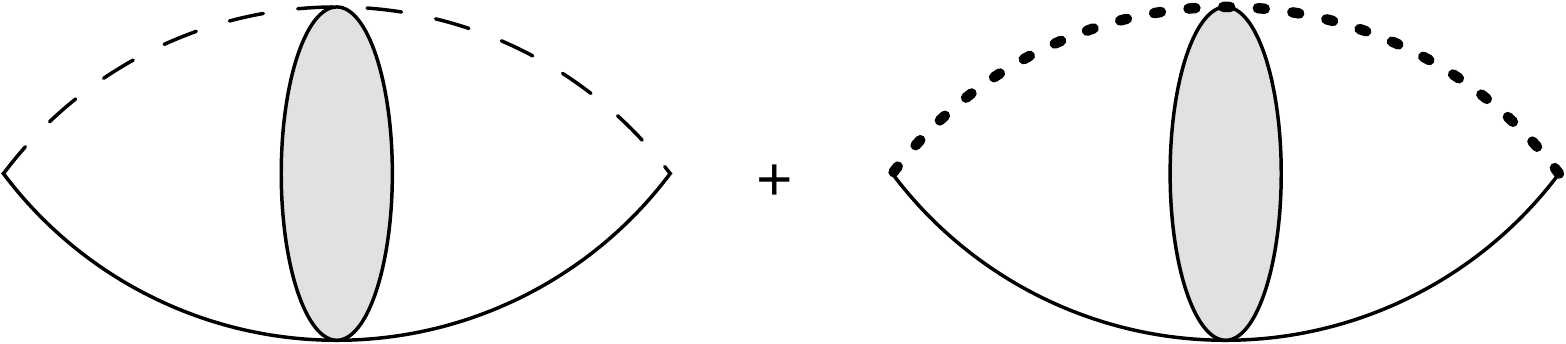}}
\caption{Irreducible self-energy loop-diagrams. The solid line
  represents the proton, the dashed line the core, and the dotted line
  the excited-core field. The shaded blob indicates the Coulomb
  Green's function.}
\label{fig:IrreducibleSelfEnergy}
\end{figure}

To describe the halo bound state, $\nuc{8}{B}$, we need the 
irreducible self-energy, which is given in terms of Feynman diagrams in
Fig.~\ref{fig:IrreducibleSelfEnergy}. There are two contributions from
(i) the core-proton bubble $\Sigma$ and (ii) the excited core-proton bubble
$\Sigma_*$.
The self energy is required for
the full form of the dressed halo propagator and the resulting
LSZ-factor for the halo bound state.
\label{sec:irred-self-energy}

The first contribution
with the ground-state core field, $\Sigma$, is proportional
to two powers of the P-wave vertex $i(g_1\mathcal{C}^\alpha_{j
  i}\mathcal{C}^{j}_{\sigma a}k_i+g_2\mathcal{C}^\alpha_{\beta
  i}\mathcal{C}^{\beta}_{\sigma a}k_i)$. As a function of
spin-state indices, the irreducible
self-energy $\Sigma^{\alpha\beta}$ is diagonal. For convenience, we therefore define 
$\Sigma=\delta^{\alpha\beta}\Sigma^{\alpha\beta}/5$, and note that
the P-wave vertex, squared and summed over spin indices evaluates to
\begin{equation}
\frac{1}{5}\Big(g_1\mathcal{C}^\alpha_{j i}\mathcal{C}^{j}_{\sigma a}k_i+g_2\mathcal{C}^\alpha_{\beta i}
\mathcal{C}^{\beta}_{\sigma a}k_i\Big)\Big(g_1\mathcal{C}^\alpha_{j' i'}\mathcal{C}^{j'}_{\sigma a}k'_{i'}+g_2
\mathcal{C}^\alpha_{\beta' i'}\mathcal{C}^{\beta'}_{\sigma a}k'_{i'}\Big)\\
=\frac{1}{3}(g_1^2+g_2^2)\kk\cdot\kk'~.
\end{equation}
We remind the reader
that for all observables considered in this paper, the couplings
$g_1$ and $g_2$ will always appear in the combination
$g^2=g_1^2+g_2^2$. 

In the following, we will require some specific 
properties of the Coulomb functions reviewed in the Appendices.
Evaluating the irreducible self-energy for the ground state core field
using Eq.~(\ref{eq:xi1b}), we find
\begin{eqnarray}
i\Sigma (E)&=&
i\frac{g^2}{3}\int\frac{\d^3k_1\d^3k_2}{(2\pi)^6}\kk_2\cdot\kk_1\langle\kk_2|G_\mrm{C}(E)|\kk_1\rangle
\nonumber\\
&=&i\frac{g^2}{3}\int\frac{\d^3p}{(2\pi)^3}\frac{{\bf
    X}(E_\pp)\cdot{\bf X}^*(E_\pp)}{E-\pp^2/(2\mR)}~.
\label{eq:Sigmaxi1}
\end{eqnarray}
The properties of the vector function
$\mbf{X}(E_\pp)=\int\d^3k~\kk\psi_\pp(\kk)/(2\pi)^3$, where $\psi_\pp$
is the Coulomb wavefunction are discussed in detail in Appendix
\ref{sec:p-wave-integrals} (cf.~Eq.~(\ref{eq:Xdef})).  Writing
$k^2=2\mR E~$ and using Eqs.~(\ref{eq:xi1b}) and (\ref{eq:CetaP}), 
we obtain the integral,
\begin{eqnarray}
i\Sigma(E) &=&-i\frac{g^2\mR}{3\pi^2}\int\d p\,\frac{C(1,\eta')^2p^4}{p^2-k^2}\nonumber\\
&=&-i\frac{g^2\mR}{3\pi^2}\int\d p\,C_{\eta'}^2\Big[p^2+k^2+\kC^2+\frac{\kC^2k^2+k^4}{p^2-k^2}\Big]\nonumber\\
&=&-ig^2\Big[L_3+(\kC^2+k^2)L_1+(\kC^2k^2+k^4)J^\mrm{fin}(k)\Big]~,
\label{eq:Sigma}
\end{eqnarray}
where we have defined
\begin{equation}
L_n=\frac{\mR}{3\pi^2}\int\d p\,C_{\eta'}^2p^{n-1}
\end{equation}
and
\begin{equation}
J^\mrm{fin}(k)=\frac{\mR}{3\pi^2}\int\d p\,\frac{C_{\eta'}^2}{p^2-k^2}~.
\end{equation}
The $L_n$ are formally infinite integrals that we will absorb in the
coupling constants $\Delta$ and $g$. Here 
we have introduced the Sommerfeld factor
$C_{\eta'}^2=2\pi\eta'/(\exp{(2\pi\eta')}-1)$, with $\eta'=\kC/p$, and the
Coulomb momentum $\kC=\Zc\alpha\mR$, which is the inverse Bohr radius
of the system, with $\Zc$ the charge number of the core.

The second contribution $\Sigma_*$, computed from the loop diagram involving
the excited core field, is given by
\begin{align}
i\Sigma_*(E)&=-ig_*^2\Big[L_3+(\kC^2+k_*^2)L_1+(\kC^2k_*^2+k_*^4)J^\mrm{fin}(k_*)\Big]~,
\label{eq:Sigmastar}
\end{align}
where $k_*=\sqrt{2 m_{\rm R}(E-E^*)}$.

We can evaluate the finite integral $J^\mathrm{fin}$ using the integral
representation of the polygamma function $\psi$ \cite{Abramowitz}.
Writing $\eta=\kC/k$, we have
\begin{eqnarray}
\psi(i\eta)+\frac{1}{2i\eta}-\log{(i\eta)}&=&-2\int_0^\infty\d\eta'\frac{\eta'}{\eta'^2-\eta^2}\frac{1}{\exp{(2\pi\eta')}-1}\\
&=&\frac{k^2}{\pi\kC}\int_0^\infty\d p\frac{1}{p^2-k^2}\frac{2\pi\kC/p}{\exp{(2\pi\kC/p)}-1}\nonumber\\
&=&\frac{3\pi k^2}{\kC\mR}J^\mrm{fin}(k)~,
\end{eqnarray}
and thus
\begin{equation}
J^\mrm{fin}(k)=\frac{\kC}{6\pi E}\Big(\psi(i\eta)+\frac{1}{2i\eta}-\log{i\eta}\Big)~.
\label{eq:Jfin}
\end{equation}
This completes the calculation of the self energy.

\subsection{Elastic proton-core scattering}
\label{sec:elastic-scattering}
\begin{figure}[t]
\centerline{
\includegraphics*[width=12cm,clip=true]{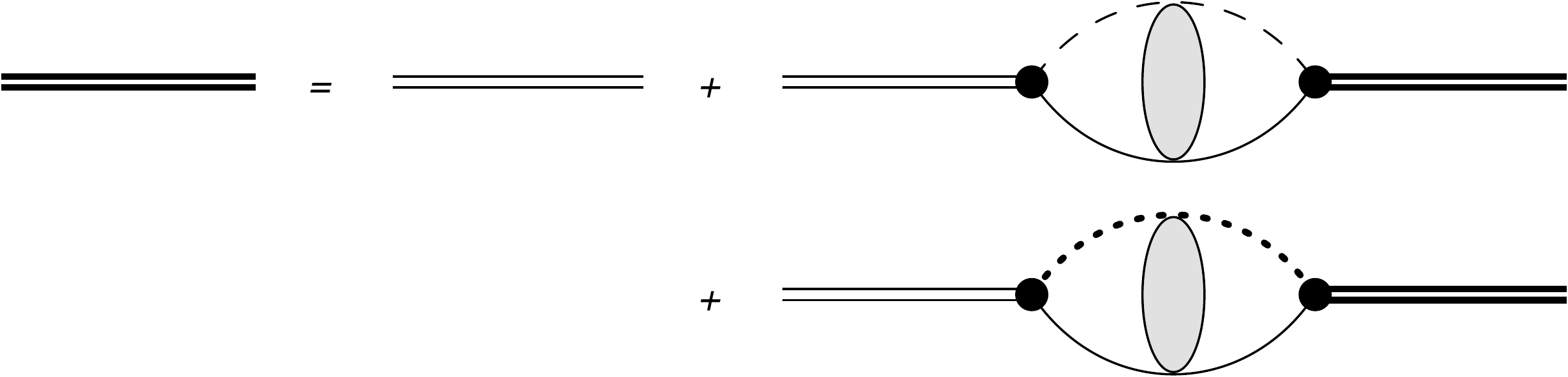}}
\caption{Integral equation for the full halo propagator. 
  The thick (thin) double line denotes the full (bare) halo propagator.}
\label{fig:Dibaryon}
\end{figure}
The low-energy coupling constants can be related to the effective-range
parameters of elastic proton-\nuc{7}{Be} scattering. To achieve this, we first
need to write down the full halo propagator with the self-energy bubble
summed to all orders.

The bare propagator $D_{\rm
  bare}$ in its center of mass frame is given by
\begin{equation}
\label{eq:bareDibProp}
iD_{\rm bare}(E)=\frac{i}{\Delta+\nu E+i\varepsilon}~,
\end{equation}
and the full propagator is given by the geometric series
\begin{eqnarray}
iD(E)
&=&iD_{\rm bare}(E)+iD_{\rm bare}(E)\, i\left[\Sigma(E)+\Sigma_*(E)\right] \, iD(E)~,
\end{eqnarray}
which is also shown schematically in Fig.~\ref{fig:Dibaryon}.  It can
be written in closed form as
\begin{align}
iD(E)&=\frac{iD_{\rm bare}(E)}{1+\left[\Sigma(E)+\Sigma_*(E) \right]D_{\rm bare}(E)}
\nonumber\\
&=\frac{i}{\Delta+\nu E+\Sigma(E)+\Sigma_*(E)+i\varepsilon}~.
\label{eq:fullDibPropnuSigma}
\end{align} 
Note that the full halo field propagator is given in terms of
irreducible self-energy bubbles from both the ground-state core field
$\Sigma$ and the excited-state core field $\Sigma_*$.

\begin{figure}[t]
\centerline{
\includegraphics*[width=4cm,clip=true]{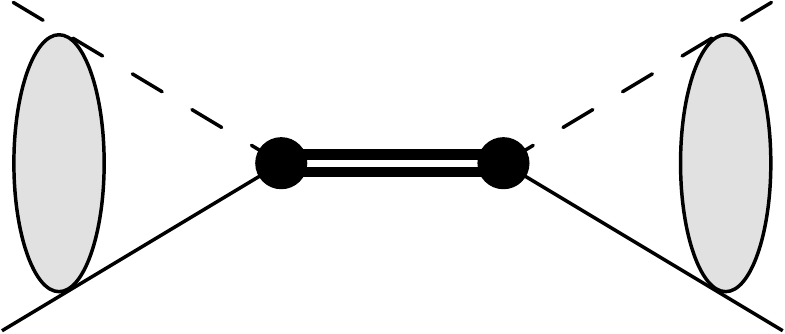}}
\caption{Elastic scattering amplitude for two charged particles with
  the Coulomb interaction included to infinite order.  The
  intermediate thick double line indicates the full halo propagator.}
\label{fig:ElasticScattering}
\end{figure}

The elastic scattering t-matrix $T_1$ is obtained by enclosing the
core-proton propagator with incoming (--) and outgoing (+) Coulomb
wavefunctions as shown in Fig.~\ref{fig:ElasticScattering}.  Using the
expressions from Eq.~(\ref{eq:xi1b}), for the P-wave integrals with the
Coulomb wavefunctions, $T_1$ is given by
\begin{align}
iT_1(E)&=ig^2D(E)~(\mbf{X}^+(E))^*\cdot\mbf{X}^-(E)\nonumber\\
&=ig^2D(E)\exp{(2i\sigma_1)}p^2C(1,\eta)^2~,
\label{eq:Tmatrix1}
\end{align}
where the P-wave pure Coulomb phaseshift is given by $\sigma_1$. The
t-matrix is given in terms of observables according to
\begin{equation}
T_1(E)=\frac{6\pi}{\mR}\frac{k^2\exp{(2i\sigma_1)}}{k^3(\cot{\delta_1}-i)}~,
\label{eq:Tmatrix2}
\end{equation}
where $\delta_1$ denotes the P-wave phaseshift due to the strong interaction.
We can then match Eqs. (\ref{eq:Tmatrix1}) and (\ref{eq:Tmatrix2}) to arrive at
\begin{align}
k^3C(1,\eta)^2(\cot{\delta_1}-i)&=\frac{6\pi}{\mR}\left[g^2D(E)\right]^{-1}~.
\label{eq:ElScatt}
\end{align}
The Coulomb-modified effective range expansion (ERE) for the P-wave
proton-$\nuc{7}{Be}$ system, with both the ground state and excited
state core fields included, is
\cite{BoGe:1984aa}
\begin{equation}
k^3C(1,\eta)^2(\cot{\delta_1}-i)+2\kC h_1(\eta)+\frac{g_*^2}{g^2}2\kC h_1(\eta_*)=-\frac{1}{a_1}+\frac{1}{2}r_1k^2+\dots~,
\label{eq:CmEREp}
\end{equation}
where $\eta_*=\kC/k_*$ and the function $h_l$ is defined as
\begin{equation}
h_l(\eta)=k^{2l}\frac{C(l,\eta)^2}{C(0,\eta)^2}\Big(\psi(i\eta)+\frac{1}{2i\eta}-\log{(i\eta)}\Big)~,
\end{equation}
with $\psi$ being the polygamma function. For $l=1$, this reduces to
\begin{equation}
h_1(\eta)=k^2(1+\eta^2)\Big(\psi(i\eta)+\frac{1}{2i\eta}-\log{(i\eta)}\Big)
=\frac{3\pi}{\mR\kC}k^4(1+\eta^2)J^\mrm{fin}(E)~,
\end{equation}
where Eq.~(\ref{eq:Jfin}) has been used in the last step.

Combining Eqs.~(\ref{eq:ElScatt}) and (\ref{eq:CmEREp}) and using
that the $J^\mrm{fin}$ and $h_1$ terms cancel out, 
we match order by order in $k^2$ to obtain
\begin{eqnarray}
a_1&=&-\frac{\mR}{6\pi}\Biggl[\frac{\Delta}{g^2}-\left(1+\frac{g_*^2}{g^2}\right)L_3
-\left(\kC^2+\kC^2\frac{g_*^2}{g^2}-2 m_{\rm R} E^*\frac{g_*^2}{g^2}\right)L_1\Biggr]^{-1}~,
\label{eq:ScattLength}\\
r_1&=&\frac{12\pi}{\mR}\Biggl[\frac{\nu}{2\mR g^2}-\left(1+\frac{g_*^2}{g^2}\right)L_1\Biggr]~.
\label{eq:EffRange1}
\end{eqnarray}
These are the P-wave scattering parameters for the $J=2$
channel. Equations~(\ref{eq:ScattLength}) and (\ref{eq:EffRange1}) define
the two renormalization conditions needed in the case of a P-wave
interaction in the presence of the Coulomb interaction.
The Coulomb-modified ERE contains the ratio of $g^2$ and $g_*^2$. We
note that these are both inversely proportional to the wavefunction
renormalization or LSZ factor $\mathcal{Z}$ of the full halo propagator
and their ratio is therefore finite.

The wavefunction renormalization  $\mathcal{Z}$ is the residue of the halo 
propagator at the bound state pole. We can thus calculate  $\mathcal{Z}$
as
\begin{eqnarray}
\mathcal{Z}&=&\left.\Big[\frac{\d(D^{-1})}{\d E}\Big]^{-1}~\right|_{E=-B}\nonumber\\
&=&\left.\Big[\nu+\Sigma'+\Sigma_*'\Big]^{-1}~\right|_{E=-B}
\label{eq:LSZnuSigma}\\
&=&\left.\frac{6\pi}{g^2\mR^2}\Biggl[r_1-\frac{2\kC}{\mR}\frac{\d}{\d E}\Biggl( h_1(\eta)+\frac{g_*^2}{g^2}h_1(\eta_*)\Biggr)\Biggr]^{-1}~\right|_{E=-B}~,
\label{eq:LSZfactor}
\end{eqnarray}
where we have used Eqs.~(\ref{eq:fullDibPropnuSigma}),
(\ref{eq:Sigma}), (\ref{eq:Sigmastar}) and (\ref{eq:EffRange1}).

\subsection{Matching to the Asymptotic Normalization Coefficients}

\begin{table}
  \caption{Asymptotic Normalization Coefficients (ANCs) calculated by
    Nollett and Wiringa \cite{Nollett:2011qf} and Zhang {\it et al.} \cite{Zhang:2014zsa},  
    by Navr\'atil {\it et al.} \cite{Navratil:2011sa},
    and extracted from a proton-transfer reaction by Tabacaru et
    al. \cite{Tabacaru:2005hv}. The ANCs are 
    given in fm$^{-1/2}$ for the two spin-channels $S=1,2$ ($A_1, A_2$)
    and for the $S=1$ 
    channel with an excited core ($A_*$).\label{table:ANCs}}
\begin{ruledtabular}
\begin{tabular}{ l c c c }
\multicolumn{1}{c}{Ref.} &$A_1$&$A_2$&$A_*$\\
\hline
``Nollett'' \cite{Nollett:2011qf}, \cite{Zhang:2014zsa} &$~-0.315(19)~$&$~-0.662(19)~$&$~0.3485(51)~$\\
``Navr\'atil'' \cite{Navratil:2011sa} &$-0.294$&$-0.650$&--\\
``Tabacaru'' \cite{Tabacaru:2005hv} &0.294(45)&0.615(45)&--\\
\end{tabular}
\end{ruledtabular}
\end{table}
Above, we derived matching conditions that relate the low-energy
coupling constants $\Delta$ and $g$, $g_*$ to the parameters of the
Coulomb-modified ERE. In practice, we require the one-proton separation
energy and the effective range as input for our calculations. However,
we can also obtain the
effective range from the Asymptotic Normalization Coefficients
(ANCs), which are defined as the coefficient of
the bound-state wavefunction outside the range $R$ of the
interaction. In this exterior region, the solution to the 
radial wavefunction $w(r)$ is
proportional to a Whittaker $W$-function in the Coulomb case. We can
therefore write
\begin{equation}
  \label{eq:anc-def}
  w_l(r)=A W_{-i\eta,l+1/2}(2\gamma r)~,\quad \hbox{for}\quad r>R~,
\end{equation}
with the binding momentum $\gamma=\sqrt{2\mR B}$ and the coefficent
$A$ defining the ANC.

The P-wave effective range for the $J=2$ channel is then obtained by
using the ANCs either from {\it ab initio} microscopic calculations,
or from a transfer reaction experiment. In this work, we compare three
different sets of input ANCs given in  Table~\ref{table:ANCs}.
Firstly, we employ ANCs from variational
Monte Carlo calculations by Nollett and Wiringa \cite{Nollett:2011qf}
and by Zhang et al. \cite{Zhang:2014zsa} for the excited-core
channel. Secondly, we employ ANCs from an NCSM/RGM calculation by
Navr\'atil et al. \cite{Navratil:2011sa}. In this case, no error
estimate was provided for the published result. Finally, we use ANCs
extracted from a proton-transfer experiment by Tabacaru et
al. \cite{Tabacaru:2005hv}.  These three sets of ANCs are denoted
``Nollett'', ``Navr\'atil'', and ``Tabacaru'', respectively, and are
listed in Table~\ref{table:ANCs}. To obtain the effective range from
the ANCs we use the relation derived by K\"onig {\it et al.} in
Ref.~\cite{Koenig:2012bv}\footnote{Here we corrected a typographical
  error in Eq.~(85) of Ref.~\cite{Koenig:2012bv}.}.  For the ground
state ANCs we obtain
\begin{equation}
A_1^2+A_2^2=2\gamma^2\Gamma{\left(2+k_\mathrm{C}/\gamma\right)}^2
\Biggl[\left.-r_1+\frac{2k_\mathrm{C}}{m_\mathrm{R}}\frac{\d}{\d
    E}\left(h_1(\eta)+\frac{g_*^2}{g^2}h_1(\eta_*)\right)\right|_{E=-B}\Biggr]^{-1}~, 
\label{eq:ANC}
\end{equation}
and for the excited state
\begin{equation}
A_*^2=2\gamma_*^2\Gamma{\left(2+k_\mathrm{C}/\gamma_*\right)}^2\Biggl[\left.-\frac{g^2}{g_*^2}r_1+\frac{2k_\mathrm{C}}{m_\mathrm{R}}\frac{\d}{\d
    E}\left(\frac{g^2}{g_*^2}h_1(\eta)+h_1(\eta_*)\right)\right|_{E=-B}\Biggr]^{-1}~, 
\label{eq:ANCstar}
\end{equation}
where $\gamma_*=\sqrt{2\mR(B+E_*)}$. The two equations \eqref{eq:ANC}
and \eqref{eq:ANCstar} determine the two unknowns $g^2/g_*^2$ and
the effective range $r_1$. Using the calculated and measured ANCs for
$^8$B in the $S=1,2$ channels in Table~\ref{table:ANCs}, we find the
effective range
\begin{equation}
r_1=\left\{\begin{array}{ll}
-(60\pm4)~\mathrm{MeV}&\mrm{(Nollett~ANCs)}\\
-63~\mrm{MeV}&\mathrm{(Navr\acute{a}til~ANCs~)}\\
-(69\pm13)~\mrm{MeV}&\mrm{(Tabacaru~ANCs)}
\end{array}\right.~,
\label{eq:effrangeANC}
\end{equation}
where we have used the excited state ANC calculated by Zhang et
al. \cite{Zhang:2014zsa} for all three results. The change in the
effective range from including the excited-core state corresponds to
an $7$-$8\%$ increase.

It is important to note that the wave function 
renormalization of the halo propagator can be expressed directly
in terms of the ANCs, according to
\begin{equation}
\mathcal{Z}=-\frac{3\pi}{g^2\mR^2\gamma^2\Gamma(2+\kC/\gamma)^2}\left(A_1^2+A_2^2\right)=
-\frac{3\pi}{g_*^2\mR^2\gamma_*^2\Gamma(2+\kC/\gamma_*)^2}A_*^2~.
\label{eq:LSZANCgs}
\end{equation}
Both expressions on the right-hand-side give the same value for the $\mathcal{Z}$-factor as one
can easily verify by using Eqs.~\eqref{eq:ANC} and
\eqref{eq:ANCstar}. Note that the LSZ residue in Eq.~\eqref{eq:LSZANCgs} is
proportional to the ground-state ANCs squared. These formulas can be
used to make predictions of bound-state properties, without having to
extract effective range parameters.

\section{The Charge Form Factor}
\label{sec:electric-form-factor}
We are now in the position to obtain results for the charge radius of
\nuc{8}{B}. The charge radius is obtained from the charge form factor
$F_\mrm{C}$, that can be measured e.g. using elastic electron-\nuc{8}{B}
scattering. The charge form factor can be extracted from the matrix element
of the zero-component of the electromagnetic current $J^\mu$
in the halo nucleus,
\begin{equation}
\langle\pp_2|J^0|\pp_1\rangle=(\Zc+1)eF_\mrm{C}(Q)~,
\label{eqDefFormFactor}
\end{equation}
evaluated in the Breit frame. The external photon momentum (or
momentum exchange) is given by $\QQ=\pp_2-\pp_1$, where $\pp_1$
($\pp_2$) is the incoming (outgoing) momentum state of the scattered nucleus.
Note that we use the standard non-relativistic
normalization of one-particle states, $\langle \pp_2|\pp_1\rangle
= (2\pi)^3 \delta^3(\pp_2-\pp_1)$.

The irreducible diagrams that we need to evaluate for this matrix element
are a tree-diagram $\Gamma_\mrm{tree}$ with the $A_0$ photon coupling
to the halo field, see Fig.~\ref{fig:GammaTree}, and a loop-diagram
$\Gamma_\mrm{loop}$ where the $A_0$ photon couples to either the 
(excited) core or the proton, see Fig.~\ref{fig:GammaLoop}.

We have calculated the charge radius at LO, using both the EFT that
contains only the ground state of the core and one that also includes
the excited state of the core. This leads to small differences in the
result that will be discussed below. Furthermore, we are limited to a
LO calculation since at NLO a short-range operator
$\mrm{d}^\dagger_\alpha\nabla^2A_0\mrm{d}_\alpha$ enters, which
requires a form factor datum for renormalization.

\subsection{Evaluation of $\Gamma_\mrm{tree}$}
\begin{figure}[t]
\centerline{
\includegraphics*[scale=0.7,clip=true]{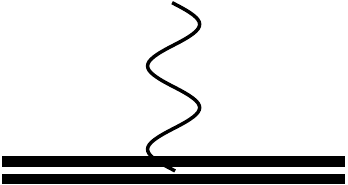}}
\caption{Feynman diagram for $\Gamma_\mrm{tree}$ diagram. The curly
  (double) line indicates the photon (halo) field.} 
\label{fig:GammaTree}
\end{figure}
The photon can couple directly to the halo field at tree level.
This diagram, which is shown in Fig. \ref{fig:GammaTree}, is given
by
\begin{equation}
i\Gamma_\mrm{tree}=i\nu e(\Zc+1)~.
\label{eq:GammaTree}
\end{equation}

\subsection{Evaluation of $\Gamma_\mrm{loop}$}
\begin{figure}[t]
\centerline{
\includegraphics*[scale=0.4,clip=true]{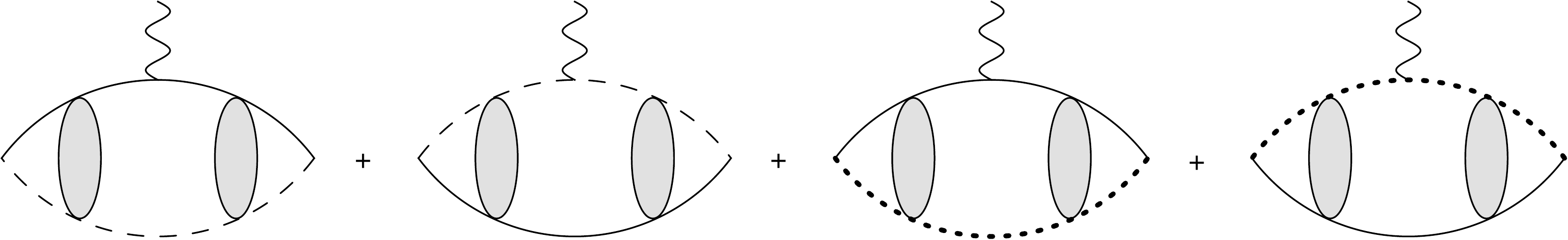}}
\caption{Feynman diagrams for $\Gamma_\mrm{loop}$ where the photon couples
  either to the proton (solid line), the core (dashed line) or excited
  core (dotted line) in the proton-core loop.}
\label{fig:GammaLoop}
\end{figure}

In total we have four loop diagrams contributing to the LO charge form
factor. These are shown in Fig.~\ref{fig:GammaLoop}. They are loop
diagrams in which an external $A_0$ photon couples either to the
proton, the core or to the excited-core field.

Let us start by writing out the integral for the ground state proton-core loop. The
momentum space integral, for momentum transfer $\QQ$, is given by
\begin{align}
\nonumber
i\Gamma_\mrm{loop,g.s.}(\QQ)=&-i\frac{g^2e\Zc}{3}\int\frac{\d^3k_1\d^3k_2\d^3k_3}{(2\pi)^{9}}\kk_3\cdot\kk_1\langle\kk_3|G_\mrm{C}(-B)|\kk_2-f\QQ/2\rangle
\\
\nonumber
&\times\langle\kk_2+f\QQ/2|G_\mrm{C}(-B)|\kk_1\rangle
\\
&+\Big[(f\to1-f),~(\Zc\to 1)\Big]~,
\label{eq:GammaLoopk}
\end{align}
where $\kk_1$ and $\kk_3$ are the loop-momenta in the outermost loops
in the diagrams in Fig.~\ref{fig:GammaLoop}, and $\kk_2$ is the
loop-momentum in the middle loop. The $\left[(f\to1-f),~(\Zc\to
  1)\right]$ term is present since the photon can couple to both the
ground state core field and to the proton. Continuing by doing a Fourier
transform on the ${\bf k}_2$ momentum, and using Eq.~(\ref{eq:I10r}),
we write this as
\begin{align}
\nonumber
i\Gamma_\mrm{loop,g.s.}(\QQ)=&-i\frac{g^2e\Zc}{3}\int\frac{\d^3k_1\d^3k_2\d^3k_3}{(2\pi)^{9}}
\d^3r_1\d^3r_2~\kk_3\cdot\kk_1\langle\kk_3|G_\mrm{C}(-B)|\rr_1) 
\\
\nonumber
&\times\exp{\left(i\kk_2\cdot(\rr_2-\rr_1)\right)}\exp{\left(if\QQ\cdot(\rr_1+\rr_2)/2\right)}(\rr_2|G_\mrm{C}(-B)|\kk_1\rangle
\\
\nonumber
&+\Big[(f\to1-f),~(\Zc\to 1)\Big]\nonumber\\
=&-i3g^2e\Zc\int\d^3r~\exp{\left(if\QQ\cdot\rr\right)}\left|\lim_{r'\to0}\left(\frac{G_\mrm{C}^{(1)}(-B;r',r)}{r'}\right)\right|^2\nonumber
\\
&+\Big[(f\to1-f),~(\Zc\to 1)\Big]~.
\end{align}
Replacing the limit of the partial-wave projected Coulomb Green's
function by Eq.~(\ref{eq:GC1W}) we arrive at
\begin{align}
\Gamma_\mrm{loop,g.s.}(Q)=&-\frac{e(\Zc+1)g^2\mR^2\Gamma{(2+\kC/\gamma)}^2\gamma^2}{3\pi}\nonumber\\
&\times\int\d r~\left[1-\Big((1-f)^2+\Zc
  f^2\Big)\frac{r^2Q^2}{6(\Zc+1)}+\Ordo(Q^4) \right]W_{-\kC/\gamma,3/2}(2\gamma
r)^2~, 
\label{eq:GammaLoopGS}
\end{align}
where $W$ is the Whittaker W-function. The order $Q^2$ integral in
Eq.~(\ref{eq:GammaLoopGS}) can be solved numerically. 

Using the same steps as above we can also derive the contribution from the
excited-core diagrams
\begin{align}
\Gamma_\mrm{loop,e.s.}(Q)=&-\frac{e(\Zc+1)g_*^2\mR^2\Gamma{(2+\kC/\gamma_*)}^2\gamma_*^2}{3\pi}\nonumber\\
&\times\int\d r~\left[1-\Big((1-f)^2+\Zc
  f^2\Big)\frac{r^2Q^2}{6(\Zc+1)}+\Ordo(Q^4)
\right]W_{-\kC/\gamma_*,3/2}(2\gamma_* r)^2~. 
\label{eq:GammaLoopES}
\end{align}

\subsection{The Charge Radius}
The charge radius $r_\mrm{C}$ is defined from the $Q^2$-term in the
expansion of the charge form factor in even powers of $Q$
\begin{equation}
\label{eq:chargeradius}
F_\mrm{C}(Q)=1-\frac{r_\mrm{C}^2}{6}Q^2+\dots\,,
\end{equation}
which is given by the sum of the three
contributions discussed above
\begin{equation}
\label{eq:ffdiag}
F_\mrm{C}(Q)=\frac{\mathcal{Z}}{e(\Zc+1)}\left(\Gamma_\mrm{tree}+\Gamma_\mrm{loop,g.s.}(Q)+\Gamma_\mrm{loop,e.s.}(Q)\right)~,
\end{equation}
with $\mathcal{Z}$ the wavefunction renormalization in Eq.~(\ref{eq:LSZfactor}).
We evaluate this expression for $Q=0$ to show that the form factor is
normalized correctly. At $Q=0$, $\Gamma_\mrm{loop,g.s.}$ in
Eq.~(\ref{eq:GammaLoopk}) simplifies to
\begin{eqnarray}
\nonumber
i\Gamma_\mrm{loop,g.s.}(0)
&=&-i\frac{g^2e(\Zc+1)}{3}\int\frac{\d^3p}{(2\pi)^3}\frac{\mbf{X}(E_\pp)\cdot\mbf{X}^*(E_\pp)}{\Big(-B-p^2/(2\mR)\Big)^2}~.
\end{eqnarray}
where we have used the $k$-space version of Eq.~(\ref{eq:CGFa}), the
orthonormality of the Coulomb wavefunctions and
Eq.~(\ref{eq:Xdef}). Now, comparison to Eq.~(\ref{eq:Sigmaxi1}) gives
\begin{equation}
\Gamma_\mrm{loop,g.s.}(0)=e(\Zc+1)\Sigma'(-B)~.
\label{eq:GammaZeroGS}
\end{equation}
For the excited core contribution we similarly have
\begin{equation}
\Gamma_\mrm{loop,e.s.}(0)=e(\Zc+1)\Sigma_*'(-B)~.
\label{eq:GammaZeroES}
\end{equation}
Equations~(\ref{eq:GammaZeroGS}) and (\ref{eq:GammaZeroES}) 
together with Eqs.~(\ref{eq:LSZnuSigma}), (\ref{eq:GammaTree}) and (\ref{eq:ffdiag})
lead to the correct normalization
\begin{equation}
F_\mrm{C}(0)
=1~.
\label{eq:normalizedFF}
\end{equation}

To obtain the charge radius we first define
\begin{equation}
\Gamma_\mrm{loop,g.s.}(Q)+\Gamma_\mrm{loop,e.s.}(Q)=\Gamma_\mrm{loop}^{(0)}
+ \Gamma_\mrm{loop}^{(2)}Q^2+\Ordo(Q^4)
\end{equation}
and use this to expand Eq.~(\ref{eq:ffdiag}) in
powers of $Q^2$
\begin{eqnarray}
F_\mrm{C}(Q)&=&\frac{\mathcal{Z}}{e(\Zc+1)}\left(\Gamma_\mrm{tree}+\Gamma_\mrm{loop}^{(0)}+\Gamma_\mrm{loop}^{(2)}Q^2\right)+\Ordo(Q^4)\nonumber\\
&=&1+\frac{\mathcal{Z}}{e(\Zc+1)}\Gamma_\mrm{loop}^{(2)}Q^2+\Ordo(Q^4)~,
\end{eqnarray}
where we used Eq.~(\ref{eq:normalizedFF}) in the last step.
The charge radius is therefore given by
\begin{equation}
r_\mrm{C}^2=-6\frac{\mathcal{Z}}{e(\Zc+1)}\Gamma_\mrm{loop}^{(2)}~.
\end{equation}
Evaluating the order $Q^2$ integrals in Eqs.~(\ref{eq:GammaLoopGS}) and
(\ref{eq:GammaLoopES}) we arrive at the values
\begin{equation}
r_\mrm{C}^2=\left\{\begin{array}{ll}
(2.56\pm0.08~\mrm{fm})^2&\mrm{(Nollett~ANCs)}\\
(2.50~\mrm{fm})^2&\mrm{(Navr\acute{a}til~ANCs)}\\
(2.41\pm0.18~\mrm{fm})^2&\mrm{(Tabacaru~ANCs)}
\end{array}\right.~,
\label{eq:chargeradiusresults}
\end{equation}
using the parameter values for the $^8$B system in Table~\ref{table:parameters}
and the ANCs of Nollett and Wiringa \cite{Nollett:2011qf}, Navr\'atil
{\it et al.} \cite{Navratil:2011sa} and Tabacaru {\it et
al.} \cite{Tabacaru:2005hv}, and the excited core ANC given in
Zhang {\it et al.} \cite{Zhang:2014zsa}.
\begin{table}
\caption{The relevant scales and parameters in the $^8$B system. See
  text for details.
\label{table:parameters}}
\begin{ruledtabular}
\begin{tabular}{c c c c c c c c }
$~\kC~$&$~\gamma~$&$~\gamma_*~$&$~r_1~$&$~k_\alpha~$&$~1/R_\mrm{\nuc{7}{Be}}~$&$~f~$&$~\Zc~$\\
\hline
$~23.79~\mrm{MeV}~$&$~14.97~\mrm{MeV}~$&$~30.39~\mrm{MeV}~$&$~\sim 60$-$70~\mrm{MeV}~$&$~50.86~\mrm{MeV}~$&$~74.55~\mrm{MeV}~$&$~1/8~$&$~4~$\\
\hline
\end{tabular}
\end{ruledtabular}
\end{table}

The errors given in Eq.~(\ref{eq:chargeradiusresults}) are due to the
uncertainties on the calculated or experimentally extracted ANCs that are
used as input.  In addition, we obtain an estimate for the LO halo EFT
error by assuming that the breakdown scale of the theory is given by
the $\alpha$-threshold of $\nuc{7}{Be}$.  It is at
$E_\alpha=1.5866~\mrm{MeV}$ \cite{Tilley:2002aa} which defines a
break-down momentum scale $k_\mrm{hi}\sim k_\alpha=
50.86~\mrm{MeV}$. The low-momentum scale is defined by the binding
momentum $k_\mrm{lo}\sim\gamma=14.97~\mrm{MeV}$. The expected EFT
error for the charge radius squared is therefore
$k_\mrm{lo}/k_\mrm{hi}\sim30\%$, which is larger than the input ANC
errors quoted above.

At the next order, the short-range operator
$\mrm{d}^\dagger_\alpha\nabla^2A_0\mrm{d}_\alpha$ enters the
electromagnetic current. The coefficient of this operator can be fit
to the charge radius or the charge form factor. We can analyze, however, the
stability of the EFT without a full higher order
calculation, by only including the minimal set of explicit degrees of
freedom at first and then add more physics by adding additional
information into the EFT.

If we remove the excited-core field from the EFT we would
arrive at the LO result $r_\mrm{C}^2=(2.32~\mrm{fm})^2$, using the
central value of the ANCs by Nollett and Wiringa \cite{Nollett:2011qf}
only.  For such a field theory, an estimate of the break-down scale
would be the momentum scale corresponding to the core excited-state
energy $\sqrt{2\mR E_*}=26.4~\mrm{MeV}$, and thus the expected EFT
error for the charge radius squared would be $57\%$. Comparing the
``Nollett ANCs'' charge radius result in
Eq.~(\ref{eq:chargeradiusresults}),
$r_\mrm{C}=(2.56\pm0.38(\mrm{EFT}))~\mrm{fm}$, with the result using
only the ground state core field
$r_\mrm{C}=(2.32\pm0.67(\mrm{EFT}))~\mrm{fm}$, we see that the size of
expected error decreases due to the inclusion of the excited state of
the $\nuc{7}{Be}$ core.  Therefore, even though we formally only
consider LO, we can test the stability of the EFT:
The inclusion of additional information at higher energies, {\it i.e.}
shorter distances, leads to a reduction in the anticipated error.

The charge radius of \nuc{8}{B} has been calculated by Pastore {\it et
al.} \cite{Pastore:2012rp} with quantum Monte Carlo techniques. The
authors find the point-proton charge radius to be $r_{\rm
  pt-p}=2.48\;\hbox{fm}$.  Folding the point-proton charge radius with
the proton and neutron charge radii, and adding relativistic plus
spin-orbit corrections as in Ref.~\cite{Ong:2010gf}, gives a charge
radius of $r_{\rm C}=2.60\;\hbox{fm}$. This value is compatible
with our LO result within the estimated error of
$30\%$ for the radius squared.

\section{Radiative capture}
\label{sec:radiative-capture}
We will now consider low-energy radiative proton capture on
\nuc{7}{Be}. We will not consider initial-wave scattering due to the
strong interaction since it is exponentially suppressed at threshold
$E_\mrm{c.m.}=0$ due to the Coulomb repulsion.  Note, however, that we
do consider pure Coulomb scattering to all orders for the incoming
proton-core pair.

\begin{figure}[t]
\centerline{
\includegraphics*[scale=0.7,clip=true]{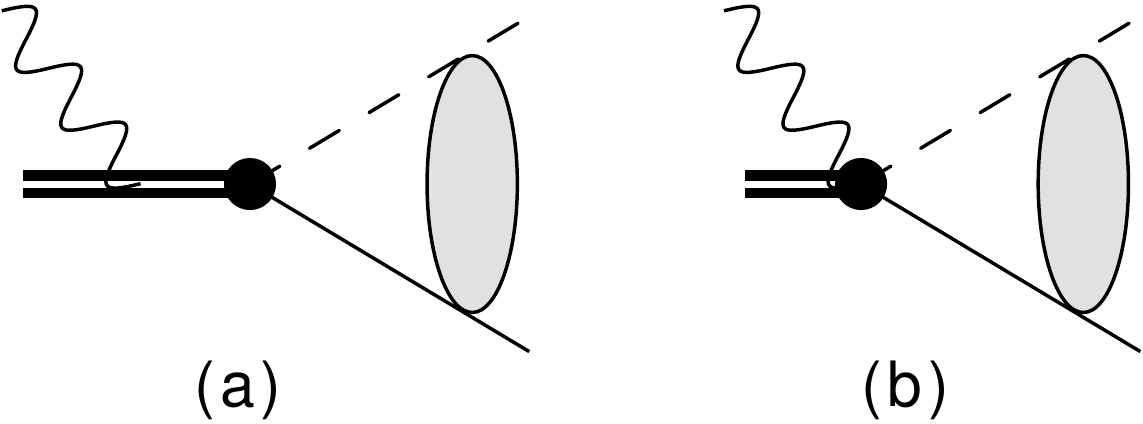}}
\caption{(a) Radiative capture diagram with an intermediate halo propagator. (b) Radiative capture diagram with a photon coupling to the P-wave vertex.}
\label{fig:CaptureAB}
\end{figure}

\begin{figure}[t]
\centerline{
\includegraphics*[scale=0.7,clip=true]{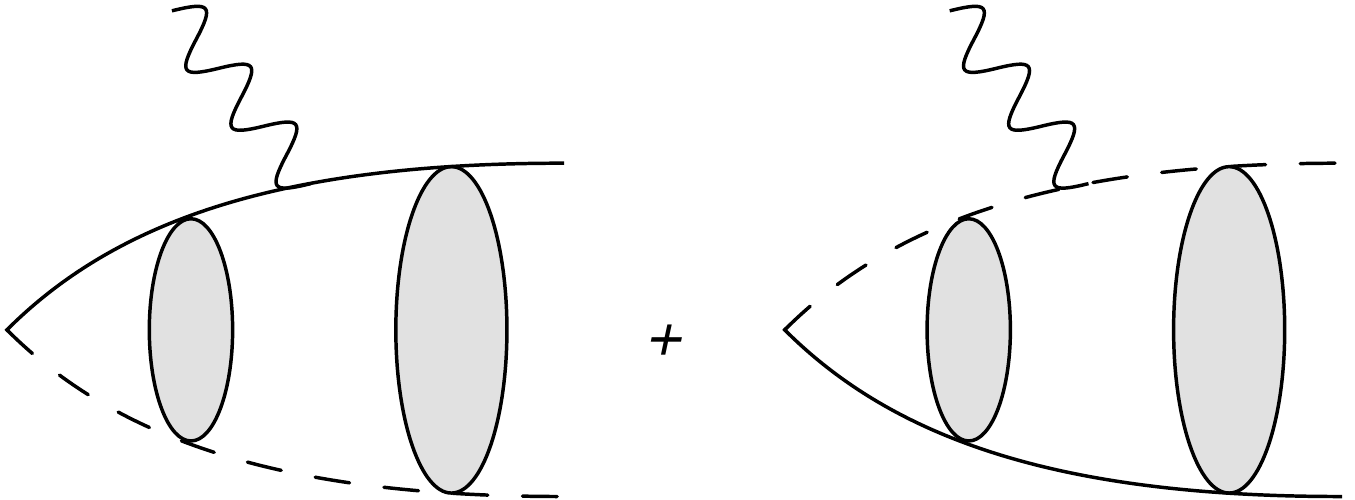}}
\caption{Loop-diagrams for radiative capture. The photon couples
  either to the proton or the core.}
\label{fig:CaptureLoop}
\end{figure}

Without initial-wave scattering we are left with the diagrams shown in
Figs.~\ref{fig:CaptureAB} and \ref{fig:CaptureLoop}, which for the
capture process has the direction of time from right to left. Below we
discuss these diagrams in detail. Since the bound state is a
$J^\pi=2^+$ due to a P-wave interaction, the E1 capture occurs through
an incoming proton-core pair in a relative S- or D-wave, with spin
channel either $S=1$ or $S=2$. These are the relevant channels since
the E1 operator changes the angular momentum by one, and the spin of
the constituent $\nuc{7}{Be}$ core is $3/2^-$ while the proton has
spin $1/2^+$. The two channels are defined by the two different
vertices in the Lagrangian Eq.~(\ref{eq:lagrangian}),
$g_1\mathcal{C}^\alpha_{k j}\mathcal{C}^k_{\sigma a}$ for the spin-1
channel ($k=-1,0,1$) and $g_2\mathcal{C}^\alpha_{\beta
  j}\mathcal{C}^\beta_{\sigma a}$ for the spin-2 channel
($\beta=-2,-1,0,1,2$), where the Clebsch-Gordan coefficients guarantee
that the spins of the constituents couple to either $S=1$ or $S=2$ and
that the total spin couples to the E1 operator to give a $J=2$ final
state. For the threshold capture cross section the S-wave dominates
and we therefore drop the D-wave contribution.

The threshold radiative capture cross section without initial wave
scattering is purely proportional to the ground state ANCs in the
combination $A_1^2+A_2^2$, since the incoming particles are in their
respective ground states. Thus, even though the excited-core field
does not contribute to the S-factor result presented here, the excited
state is included in the field theory and we may therefore estimate
the error from the $\alpha$-breakup threshold of $\nuc{7}{Be}$. The
results we present are at LO, but since the pole-position is fixed by
the ANCs, the threshold S-factor should be accurate. This is
because the higher-order operators in the field theory will enter with
additional powers of the energy. Therefore the error for the threshold
S-factor is mainly due to the uncertainty in the input ANCs, while at
finite energies the error is from neglected initial-wave
interactions, D-wave component and higher-order operators.

The amplitudes presented in this section will have various indices,
one from the vector photon $A^i$ and three from the incoming and
outgoing particles $(\alpha,\sigma,a)$. The first index will thus be
dotted by the polarization vectors and the remaining (written in
parenthesis) will be traced in the summation of $|\mathcal{A}|^2$.
Furthermore, the diagrams are evaluated in the zero-momentum frame of
the proton and the core. The incoming proton (core) has thus momentum
${\bf p}$ ($-{\bf p}$). The outgoing photon (halo) has momentum ${\bf
  Q}$ ($-{\bf Q}$).

\paragraph{Capture diagram with intermediate halo propagator}

This is the simplest diagram, and is shown in
Fig. \ref{fig:CaptureAB}(a). Since the amplitude is proportional to
the momentum flowing into the photon-halo field vertex, in Coulomb
gauge, then the diagram is identically zero in the zero-momentum frame
\begin{equation}
i\mathcal{A}^{i(\alpha)}_{1(\sigma a)}=0~.
\label{eq:capture1}
\end{equation}

\paragraph{Capture diagram with the photon coupling to the P-wave vertex}

This diagram is given in Fig. \ref{fig:CaptureAB}(b) and is simply 
\begin{align}
  i\mathcal{A}_{2(\sigma a)}^{i(\alpha)}=&-ie\Big(1-f(\Zc+1)\Big)
\left(g_1 \mathcal{C}^\alpha_{j i}\mathcal{C}^j_{\sigma a}+g_2\mathcal{C}^\alpha_{\beta i}\mathcal{C}^\beta_{\sigma a}\right)\int\frac{\d^3k}{(2\pi^3)}~\psi_\pp(\kk)\nonumber\\
  =&-ie\Big(1-f(\Zc+1)\Big)\left(g_1 \mathcal{C}^\alpha_{j
      i}\mathcal{C}^j_{\sigma a}+g_2 \mathcal{C}^\alpha_{\beta i} \mathcal{C}^\beta_{\sigma a}\right)\psi_\pp(0)~.
\label{eq:capture2}
\end{align}
The $\psi_\pp(0)$ factor implies that this amplitude is purely from an incoming S-wave.

\paragraph{Capture loop diagrams}

The loop diagrams contributing to the capture process are shown in
Fig. \ref{fig:CaptureLoop} and are written in momentum space as
\begin{eqnarray}
i\mathcal{A}^{i(\alpha)}_{3(\sigma a)}&=&\left(g_1
  \mathcal{C}^\alpha_{k j}\mathcal{C}^k_{\sigma a}
+g_2\mathcal{C}^\alpha_{\beta j}\mathcal{C}^\beta_{\sigma a}\right)\int\frac{\d^3k_1\d^3k_2}{(2\pi)^6}~ik_2^j\langle\kk_2|G_\mrm{C}(-B)|\kk_1+f\QQ\rangle i\frac{fe\Zc(-k_1^i)}{\mR}\psi_\pp(\kk_1)\nonumber\\
&&-\left[(f\to1-f)~,~(\Zc\to1)\right]~.
\end{eqnarray}
It involves an incoming Coulomb wavefunction $\psi_\pp(\kk_1)$, the
vector photon-core vertex $ife\Zc(-k_1^i)/\mR$, a Coulomb Green's
function $G_\mrm{C}(-B)$, which defines the propagation down to the
bound state, and the P-wave interaction vertex
$ik_2^j\left(g_1\mathcal{C}^\alpha_{k j}\mathcal{C}^k_{\sigma
    a}+g_2\mathcal{C}^\alpha_{\beta j}\mathcal{C}^\beta_{\sigma
    a}\right)$. The photon-proton vertex is given by
$i(1-f)ek_1^i/\mR$. Fourier transforming to coordinate space and
multiplying with the photon polarization vectors $\epsilon_i^{(n)}$,
where $n=1,2$ and we have chosen $\epsilon^{(1)}=\hat{x}$,
$\epsilon^{(2)}=\hat{y}$ and $\QQ=\omega\hat{z}$, we arrive at
\begin{eqnarray}
\sum_{i=-1}^1\sum_{n=1}^2i\epsilon_i^{(n)}\mathcal{A}_{3(\sigma a)}^{i(\alpha)}&=&-\left(g_1\mathcal{C}^\alpha_{k j}\mathcal{C}^k_{\sigma a}
+g_2\mathcal{C}^\alpha_{\beta j}\mathcal{C}^\beta_{\sigma a}\right)
\frac{2fe\Zc\gamma}{3p}(\hat{x}+\hat{y})^j\exp{(i\sigma_{0})}\Gamma{(2+\kC/\gamma)}\nonumber\\
&&\times\int\d r~rW_{-\kC/\gamma,3/2}(2\gamma r) j_{0}(f\omega r)\partial_r\left(\frac{F_{0}(\kC/p,pr)}{r}\right)\nonumber\\
&&-\left[(f\to1-f)~,~(\Zc\to1)\right]~,
\label{eq:capture3}
\end{eqnarray}
where we have dropped the incoming D-wave component and a negligible $j_2$
spherical Bessel function.

\paragraph{Resulting S-factor}
The LO cross section for radiative capture is now given by the sum of
contributions discussed above
\begin{equation}
\label{eq:diffcross}
\frac{\d\sigma}{\d\Omega}=
\frac{\mR\omega}{8\pi^2p}\sum_{\alpha,\sigma
  a}\sum_{i=-1}^1\sum_{n=1}^2\bigg|\sqrt{\mathcal{Z}}\epsilon_i^{(n)}
\Big(\mathcal{A}^{i(\alpha)}_{1(\sigma a)}+\mathcal{A}^{i(\alpha)}_{2(\sigma a)}+\mathcal{A}^{i(\alpha)}_{3(\sigma a)}\Big)\bigg|^2~,
\end{equation}
using the expressions for the amplitudes in Eqs.~(\ref{eq:capture1}),
(\ref{eq:capture2}) and (\ref{eq:capture3}). Note that the amplitudes in
Eq.~\eqref{eq:diffcross} do not depend on the proton-core scattering
effective range. This quantity, which was determined through the ANCs,
enters solely through the wavefunction renormalization $\mathcal{Z}$
defined in Eq.~\eqref{eq:LSZfactor}.

We present the total
cross section in terms of the astrophysical S-factor
\begin{equation}
S(E)=E\exp{(2\pi\eta)}\sigma_\mrm{tot}~,
\end{equation}
which is basically defined by the removal of the low-energy
exponential suppression $C_\eta^2$.

\begin{figure}[t]
\centerline{
\includegraphics*[width=10cm,clip=true]{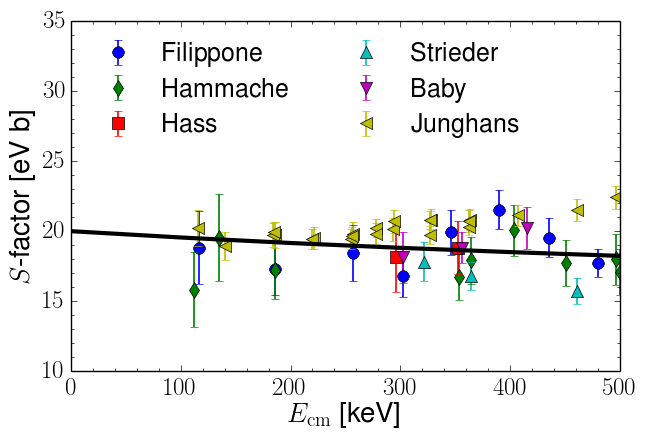}}
\caption{The S-factor of $\nuc{7}{Be}(p,\gamma)\nuc{8}{B}$ as a
  function of the c.m. energy. The 
  data are from Refs.~\cite{Filippone:1983zz, Hammache:1997rz, Hammache:2001tg, Hass:1999sp,
Strieder:2001, Baby:2002hj, Junghans:2003bd, Junghans:2010zz}. The solid
line is the LO result of this work using input ANCs of Nollett {\it et al.}.
\label{fig:Sfactor}}
\end{figure}

The resulting S-factor is presented in Fig.~\ref{fig:Sfactor} together
with existing low-energy direct-capture data from Refs.~\cite{Filippone:1983zz,
  Hammache:1997rz, Hammache:2001tg, Hass:1999sp, Strieder:2001,
  Baby:2002hj, Junghans:2003bd, Junghans:2010zz}.
The threshold value is
\begin{equation}
S(0)=\left\{\begin{array}{ll}
(20.0\pm1.4)~\mathrm{eV~b}&\mrm{(Nollett~ANCs)}\\
18.9~\mathrm{eV~b}&\mrm{(Navr\acute{a}til~ANCs)}\\
(17.3\pm3.0)~\mathrm{eV~b}&\mrm{(Tabacaru~ANCs)}
\end{array}\right.~.
\label{eq:ThresholdSFactorResult}
\end{equation}
The errors given in Eq.~(\ref{eq:ThresholdSFactorResult}) are due to
the errors in the input ANCs.  It is important to note that in the capture
process the intrinsic EFT error is smaller than the naively expected one. This matrix
element is insensitive to uncertainties in the core-proton
P-wave interaction since the incoming core and proton are not in a relative
P-wave. Furthermore, the binding energy of the final state is fixed
and the Z-factor is directly proportional to the ANCs squared as showed
in Eq. \eqref{eq:LSZANCgs}. Thus, at threshold the only corrections are due to
higher-order operators that involve the photon field and these
enter with additional powers of the photon energy $\omega$. Momentum
and energy conservation implies that $\omega \approx B+{\textstyle
  \frac{p^2}{2m_{\rm R}}}$, where the binding energy scales as $k_{\rm
  lo}^2$. We therefore estimate the error to be of order $(k_{\rm
  lo}/k_{\rm hi})^2\approx 8 \%$.
These higher-order corrections will, however, influence the shape of
the S-factor at larger energies, {\it i.e.} in the region where comparison
with data is possible. This results in an additional uncertainty when
attempting extrapolation to threshold energies. Note that this
extrapolation uncertainty is much more serious in potential-model
descriptions for which the error is not even quantified.
Therefore, we present an alternative approach to constrain the
threshold S-factor from experimental data: namely to identify its
correlation with the previously discussed charge radius of \nuc{8}{B}.

\section{Correlating the Charge Radius and the Threshold S-factor}
\label{sec:corr-charge-radi}
\begin{figure}[t]
\centerline{
\includegraphics*[width=10cm,clip=true]{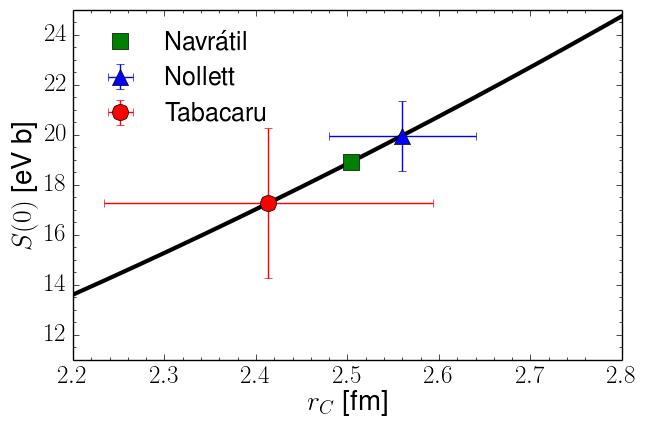}}
\caption{Correlating the threshold S-factor and the charge radius of
  \nuc{8}{B}. The solid line demonstrates the EFT correlation, while
  the three data points with error bars correspond to results obtained
  with different input ANCs, either from microscopic ab initio
  calculations, or from a transfer experiment.}
\label{fig:CorrelatingSandR}
\end{figure}

At this stage we are able to demonstrate the relationship between the
$\nuc{7}{Be}(p,\gamma)\nuc{8}{B}$ S-factor at threshold and the
\nuc{8}{B} charge radius. Instead of using ground-state ANCs from
microscopic calculations or transfer experiments as input, we now let
$A_1^2+A_2^2$ be a free parameter. This parameter is then used to
explore correlations between the charge radius of $^8$B and the
threshold S-factor.  However, we still fit the excited state ANC to
the result of Zhang {\it et al.} $A_*=0.3485(51)$. The resulting
correlation plot is shown in Fig.~\ref{fig:CorrelatingSandR}. The
solid line shows the LO one-parameter correlation predicted by halo
EFT. The triangle, square and circle show our results for the S-factor
with the input ANCs from Nollett, Navr\'atil and Tabacaru,
respectively. They appear in different positions on the correlation
line since changing the ANC leads to a different effective range and
thus a different charge radius. In turn, the correlation can be used
to constrain the ANC from charge radius measurements.

\section{Summary}
\label{sec:summary}
In this paper, we have calculated the charge radius and the S-factor
for radiative proton capture on $^7$Be at LO in halo EFT. The
parameters required to fix the proton-core interaction were obtained
from experiment, or from ab initio calculations. Our prediction for
the S-factor at threshold is in agreement with the most recent
recommended value for this observable~\cite{Adelberger:2010qa} and the
EFT calculation by Zhang, Nollett and Phillips~\cite{Zhang:2014zsa}.
By performing calculations with and without the excited state of
\nuc{7}{Be}, and thereby changing the amount of microscopic physics
included in the EFT, we have tested the stability of our
calculation. Furthermore, we have derived a number of universal
expressions that can be applied to any proton halo system bound
through resonant p-wave interactions.

If the one-proton separation energy of $^8$B is kept fixed, the charge
radius is directly correlated to the S-factor at threshold at this
order in the calculation. A measurement of the S-factor therefore
determines not only the ANCs of the \nuc{7}{Be}-proton system but also
the charge radius. This result provides an excellent example how
different observables constrain each other. Note however, that this
direct correlation does not exist at the next order in the EFT
expansion since two additional counterterms enter.  The first
counterterm fixes the next contribution in the ERE of
\nuc{7}{Be}-proton scattering. The second counterterm comes from an
operator that couples the photon to the full halo field. Both
counterterms could be fitted to the charge form factor, either
obtained through a measurement or from an ab initio calculation. This
connection to the charge form factor is particularly useful for the
latter case since calculations of {\it static} nuclear properties are
generally simpler than reactions or scattering observables.
A higher order analysis would thus provide additional
constraints on the threshold S-factor and does not directly depend on a
measurement of the capture cross section and a subsequent extrapolation.

It is worth noting that the additional counterterm appearing in the
electromagnetic current is an inherent limitation for any cluster model
that tries to describe charge radii as it accounts for microscopic
physics not determined by core-proton scattering properties. Previous
calculations using cluster models have tried to circumvent this
limitation by taking additional microscopic physics such as core
swelling into account, see e.g.~\cite{2011PhRvC..84e1304P}. However, the
model-dependence of such approaches has not been studied and involves
uncontrolled errors. Our EFT approach can therefore provide lower error
bounds on any cluster model that employs the same number of degrees of
freedom.

\begin{acknowledgments}
  We thank D.~R. Phillips and H.~Esbensen for useful discussions.  This
  research was supported in part by the Swedish Research Council
  (dnr. 2010-4078), the European Research Council under the European
  Community's Seventh Framework Programme (FP7/2007-2013) / ERC grant
  agreement no.~240603, the BMBF under grant 05P12PDFTE, 
  the DFG through SFB 634, the Office of
  Nuclear Physics, U.S.~Department of Energy under contract
  no. DE-AC02-06CH11357 and by the Helmholtz Association under contract
  HA216/EMMI.  We express our appreciation to the Extreme Matter
  Institute at GSI and the Institute for Nuclear Theory in Seattle,
  where part of this work was carried out.
\end{acknowledgments}

\begin{appendix}
\section{Coulomb wavefunctions and Green's function}
\label{app:Coulomb}

We define the Coulomb wavefunction through its partial wave
expansion
\begin{equation}
\psi_\kk(\rr)=\sum_{l=0}^{\infty}\phi_l(\eta,\rho)P_l(\hat{\kk}\cdot\hat{\rr})~,
\end{equation}
with
\begin{equation}
\phi_l(\eta,\rho)=(2l+1)i^l\exp{(i\sigma_l(\eta))}\frac{F_l(\eta,\rho)}{\rho}~,
\end{equation}
and the Coulomb phase shift $\sigma_l=\arg{\Gamma(l+1+i\eta)}$~.
Here we have used the definitions 
$\rho=kr~$ and $\eta=\kC/k~,$ with the Coulomb momentum
$\kC=\Zc\alpha\mR$~. The regular Coulomb wavefunctions
$F_l$ can be expressed in terms of the Whittaker M function through
\begin{equation}
F_l(\eta,\rho)=A(l,\eta)M_{i\eta,l+1/2}(2i\rho)~,
\end{equation}
with
\begin{equation}
A(l,\eta)=\frac{1}{2}\frac{\Gamma{(l+1+i\eta)}\exp{\Big(-\pi\eta/2-i(l+1)\pi/2\Big)}}{(2l+1)!}~.
\end{equation}
We also give the irregular Coulomb wavefunctions $G_l$ in terms of the Whittaker M and W functions as
\begin{equation}
G_l(\eta,\rho)=iF_l(\eta,\rho)+B_l(\eta)W_{i\eta,l+1/2}(2i\rho)~,
\end{equation}
with
\begin{equation}
B_l(\eta)=\frac{\exp{(\pi\eta/2+il\pi/2)}}{\arg{\Gamma{(l+1+i\eta)}}}~.
\end{equation}
Moreover, we have defined
\begin{equation}
\left|\Gamma{(l+1+i\eta)}\right|^2=\Gamma{(l+1+i\eta)}\Gamma{(l+1-i\eta)}
\end{equation}
and 
\begin{equation}
\arg{\Gamma{(l+1+i\eta)}}=\sqrt{\frac{\Gamma{(l+1+i\eta)}}
{\Gamma{(l+1-i\eta)}}}~,
\end{equation}
such that our expressions make sense for imaginary momentum
corresponding to bound states.

The Sommerfeld factor is given by
\begin{eqnarray}
C^2_\eta=C(0,\eta)^2&=&\frac{2\pi\eta}{\exp{(2\pi\eta)}-1}
=\exp{(-\pi\eta)}\Gamma{(1+i\eta)}\Gamma{(1-i\eta)}~,
\end{eqnarray}
and we will also be needing its generalization to higher partial waves
\begin{equation}
C(l,\eta)^2=\exp{(-\pi\eta)}\Gamma{(l+1+i\eta)}\Gamma{(l+1-i\eta)}~.
\end{equation}
Note in particular that 
\begin{equation}
C(1,\eta)^2=(1+\eta^2)C_\eta^2~.
\label{eq:CetaP}
\end{equation}

The partial wave projected Coulomb Green's function for a bound state is given by
\begin{eqnarray}
G_\mrm{C}^{(l)}(E;r,r')&=& -\frac{\mR p}{2\pi}\frac{F_l(\eta,\rho')\Big[iF_l(\eta,\rho)+G_l(\eta,\rho)\Big]}{\rho'\rho}~,
\label{eq:CoulGreensL}
\end{eqnarray}
that is a product of two linear combinations of Coulomb wavefunctions
satisfying the bound state boundary conditions both at zero and
infinite separation. The normalization of the Coulomb Green's function
Eq.~(\ref{eq:CoulGreensL}) is set by the Coulomb-Schr\"odinger
equation and the Wronskian of the Coulomb wavefunctions.

Using the identity
\begin{equation}
iF_l(\eta,\rho)+G_l(\eta,\rho)=\exp{(i\sigma_l+\pi\eta/2-li\pi/2)}W_{-i\eta,l+1/2}(-2i\rho)
\end{equation}
and the limit
\begin{equation}
\lim_{r\to0}\left(\frac{F_1(\eta,\rho)}{\rho^{2}}\right)= \frac{1}{3}\exp{(-\pi\eta/2)}\sqrt{\Gamma{(2+i\eta)}\Gamma{(2-i\eta)}}~,
\end{equation}
we can then  write
\begin{equation}
\lim_{\rho'\to0}\left(\frac{G_\mrm{C}^{(1)}(E;r',r)}{\rho'}\right)=i\frac{\mR p}{6\pi}\Gamma{(2+i\eta)}\frac{W_{-i\eta,3/2}(-2i\rho)}{\rho}~.
\label{eq:GC1W}
\end{equation}
The expression Eq.~(\ref{eq:GC1W}) is used to evaluate
the loop-integrals for the charge form factor and the radiative 
capture cross section.

\section{P-wave Integrals}
\label{sec:p-wave-integrals}
When we calculate loop diagrams with P-wave interactions, we
have to evaluate integrals that involve Coulomb wavefunctions. 
In this Appendix, we derive a few useful identities for these
calculations.

The most basic constituent of the integrals that we need to solve is
the vector integral
\begin{equation}
{\bf X}(E_\pp)=\int\frac{\d^3k}{(2\pi)^3}\kk\psi_\pp(\kk)~.
\label{eq:Xdef}
\end{equation}
We will evaluate this integral by performing a Fourier transform
of $\psi_\pp(\kk)$ to position space
followed by a partial integration:
\begin{eqnarray}
{\bf X}(E_\pp)&=&\int\frac{\d^3k}{(2\pi)^3}\kk\psi_\pp(\kk)\nonumber
=\int\d^3r\psi_\pp(\rr)(-i\nabla){\delta^{(3)}(\rr)}\nonumber
\\
&=&i\int\d^3r\delta^{(3)}(\rr)\Big(\nabla\psi_\pp(\rr)\Big)~.
\label{eq:xi1a}
\end{eqnarray} 
We will now temporarily fix $\pp=p\hat{z}$ and do a partial wave
expansion of the Coulomb wavefunction. The derivative in
Eq.~(\ref{eq:xi1a}) is then evaluated to (we are suppressing the
$\eta'$ and $\rho$ dependencies)
\begin{equation}
\nabla\psi_\pp(\rr)=\sum_{l=0}^\infty\Big[\hat{r}p\partial_\rho\phi_lP_l(\cos{\theta})+\hat{\theta}\frac{p}{\rho}\phi_l\partial_\theta P_l(\cos\theta)\Big]
\end{equation}
Note that for $l=0$ we get 
\begin{equation}
\nabla \psi_\pp(\rr)\Big|_{l=0}=\hat{r}p\partial_\rho\phi_0~,
\end{equation}
which is an odd function of $\rr$. This means that the integration of
the $l=0$ term is zero. Furthermore, we have
\begin{equation}
\lim_{\rho\to0}\partial_\rho\phi_l=0~,~l>1~,
\end{equation}
and
\begin{equation}
\lim_{\rho\to0}\frac{\phi_l}{\rho}=0~,~l>1~,
\end{equation}
which means that only the $l=1$ term will contribute to ${\bf X}$:
\begin{eqnarray}
{\bf X}(E_\pp)
&=&i\lim_{\rho\to0}\Big(\cos{(\theta)}\hat{r}p\partial_\rho\phi_1-\sin{(\theta)}\hat{\theta}p\frac{\phi_1}{\rho}\Big)\nonumber\\
&=&i\pp\lim_{\rho\to0}\Big(\frac{\phi_1(\eta',\rho)}{\rho}\Big)~.
\label{eq:xi1ab}
\end{eqnarray}
In the last step we used that
\begin{equation}
\cos{(\theta)}\hat{r}-\sin{(\theta)}\hat{\theta}=\hat{z}
\end{equation}
and that the limits of $\partial_\rho\phi_1$ and $\phi_1/\rho$ can be written as
\begin{equation}
\lim_{\rho\to0}\left(\partial_\rho\phi_1(\eta',\rho)\right)=\lim_{\rho\to0}\Big(\frac{\phi_1(\eta',\rho)}{\rho}\Big)~.
\end{equation}
We can also write this limit using the Sommerfeld factor
\begin{equation}
\lim_{\rho\to0}\Big(\frac{\phi_1(\eta',\rho)}{\rho}\Big)=i\exp{(i\sigma_1)}C(1,\eta')~,
\label{eq:limphi}
\end{equation}
that is
\begin{equation}
{\bf X}(E_\pp)=-\pp\exp{(i\sigma_1)}C(1,\eta')~.
\label{eq:xi1b}
\end{equation} 

\section{Partial Wave decomposed Coulomb Green's Function}

We now continue by analyzing the Coulomb Green's function
$(\rr_1|G_\mrm{C}|\rr_2)$. It is useful to express the Green's function in its partial wave
expanded form
\begin{equation}
(\rr_1|G_\mrm{C}(E)|\rr_2)=\sum_{l=0}^\infty (2l+1)G_\mrm{C}^{(l)}(E;r_1,r_2)P_l(\hat{\rr}_1\cdot\hat{\rr}_2)~.
\end{equation}
The form of $G_\mrm{C}^{(l)}$ is derived by expanding the Coulomb
wavefunctions in Eq.~(\ref{eq:CGFa}), using spherical
harmonics. In the first step we use the orthogonality of the harmonics
and in the second step the addition theorem is used:
\begin{eqnarray}
(\rr_1|G_\mrm{C}(E)|\rr_2)&=&
\sum_{l_1m_1}\sum_{l_2m_2}\int\frac{\d^3p}{(2\pi)^3}
\frac{(4\pi)^2}{(2l_1+1)(2l_2+1)}
\frac{\phi_{l_1}(\eta,\rho_1)\phi_{l_2}^*(\eta,\rho_2)}{E-\frac{\pp^2}{2\mR}}\nonumber\\
&&\quad \times Y_{l_1m_1}(\theta_1,\varphi_1)Y^*_{l_1m_1}(\theta_p,\varphi_p)Y_{l_2m_2}(\theta_p,\varphi_p)Y^*_{l_2m_2}(\theta_2,\varphi_2)\nonumber\\
&=&\sum_{lm}\int\frac{\d^3p}{(2\pi)^3}\frac{4\pi}{(2l+1)^2}\frac{\phi_{l}(\eta,\rho_1)\phi_{l}^*(\eta,\rho_2)}{E-\frac{\pp^2}{2\mR}}
\times Y_{lm}(\theta_1,\varphi_1)Y^*_{lm}(\theta_2,\varphi_2)\nonumber\\
&=&\sum_{l}\frac{P_l(\hat{\rr}_1\cdot\hat{\rr}_2)}{2l+1}\int\frac{\d^3p}{(2\pi)^3}\frac{\phi_{l}(\eta,\rho_1)\phi_{l}^*(\eta,\rho_2)}{E-\frac{\pp^2}{2\mR}}~.
\end{eqnarray}
Thus, the Green's function for a specific partial wave is given by
\begin{equation}
(2l+1)G_\mrm{C}^{(l)}(E;r_1,r_2)=\frac{1}{2l+1}\int\frac{\d^3p}{(2\pi)^3}\frac{\phi_{l}(\eta,\rho_1)\phi_{l}^*(\eta,\rho_2)}{E-\frac{\pp^2}{2\mR}}~.
\label{eq:scalarCGf}
\end{equation}

We are now in a position to simplify the integral
\begin{equation}
I_1(0,\rr)=\int\frac{\d^3k}{(2\pi)^3}\kk\langle\kk|G_\mrm{C}(E)|\rr)~.
\end{equation}
Note that it is defined by a mixed matrix element where the Coulomb
Green's function is evaluated between a bra in $k$-space and a ket
in $r$-space. Comparing to the S-wave equivalent
\begin{eqnarray}
I_0(0,\rr)&=&\int\frac{\d^3k}{(2\pi)^3}\langle\kk|G_\mrm{C}(E)|\rr)\nonumber\\
&=&(0|G_\mrm{C}(E)|\rr)~,
\end{eqnarray}
we expect that $I_1$ behaves as an $r$-space object.  
Using Eq.~(\ref{eq:xi1ab}) and doing a partial wave
expansion of $\psi^*_\pp(\rr)$, we have
\begin{eqnarray}
I_1(0,\rr)&=&\int\frac{\d^3k\d^3p}{(2\pi)^6}\kk\,\psi_\pp(\kk)\frac{\psi_\pp^*(\rr)}{E-\frac{\pp^2}{2\mR}}\nonumber\\
&=&\sum_l\int\frac{p^2\d p}{(2\pi)^3}i\lim_{\rho'\to0}\Big(\frac{\phi_1(\eta,\rho')}{\rho'}\Big)\frac{\phi_l^*(\eta,\rho)}{E-\frac{\pp^2}{2\mR}}\int\d\Omega \pp P_l(\cos{\theta})\nonumber\\
&=&i\frac{\hat{\rr}}{3}\lim_{\rr'\to0}\Big[\frac{1}{r'}\int\frac{\d^3p}{(2\pi)^3}\frac{\phi_1(\eta,\rho')\phi_1^*(\eta,\rho)}{E-\frac{\pp^2}{2\mR}}\Big]~.
\end{eqnarray}
Finally, using Eq. (\ref{eq:scalarCGf}) we find that
\begin{eqnarray}
I_1(0,\rr)&=&\int\frac{\d^3k}{(2\pi)^3}\kk\langle\kk|G_\mrm{C}(E)|\rr)\nonumber\\
&=&3i\hat{\rr}p\lim_{\rr'\to0}\Big(\frac{G_\mrm{C}^{(1)}(E;r',r)}{\rho'}\Big)~.
\label{eq:I10r}
\end{eqnarray}
This form is useful since the partial wave projected Green's function can
be written in a closed functional form. Eq.~(\ref{eq:I10r}) can be further
simplified using Eq.~(\ref{eq:GC1W}).

\end{appendix}

\end{document}